\documentclass[reprint,
nofootinbib,
 amsmath,amssymb,
 aps,
showpacs
]{revtex4-2}

\usepackage{graphicx}
\usepackage{dcolumn}
\usepackage{bm}

\usepackage[colorlinks]{hyperref}
\hypersetup{linkcolor=red,citecolor=blue,urlcolor=blue}

\begin{document}

\title{{Surface and curvature properties} of charged strangelets in compact objects}

\author{G. Lugones$^{1}$ and A. G. Grunfeld$^{2,3}$}

\affiliation{$^1$ Universidade Federal do ABC, Centro de Ci\^encias Naturais e Humanas, Avenida dos Estados 5001- Bang\'u, CEP 09210-580, Santo Andr\'e, SP, Brazil.\\
$^2$CONICET, Godoy Cruz 2290, Buenos Aires, Argentina \\
$^3$Departamento de F\'\i sica, Comisi\'on Nacional de Energ\'{\i}a At\'omica, Av. Libertador 8250, (1429) Buenos Aires, Argentina}

\begin{abstract}
Droplets of absolutely stable strange quark matter (strangelets) immersed in a lepton background may be the energetically preferred composition of strange star crusts and may also constitute the interior of a new class of stellar objects known as strangelet dwarfs.  In this work we calculate the surface tension $\sigma$ {and the curvature coefficient $\gamma$} of charged strangelets as a function of the baryon number density, the temperature, the chemical potential of trapped neutrinos, the strangelet size, the electric potential and the electric charge at their boundary. Strange quark matter in chemical equilibrium and with global electric charge neutrality is described within the MIT bag model.  We focus on three different astrophysical scenarios, namely cold strange stars, proto strange stars and post merger strange stars,   characterized by the temperature and the amount of neutrinos trapped in the system. Finite size effects are implemented within the multiple reflection expansion framework. 
We find that $\sigma$ decreases significantly  as the strangelet's boundary becomes more positively charged.  This occurs because $\sigma$ is dominated by the contribution of $s$ quarks which are the most massive particles in the system. Negatively charged $s$-quarks are suppressed in strangelets with a large positive electric charge,  diminishing their contribution to $\sigma$ and resulting in smaller values of the total surface tension.  We also verify that the more extreme astrophysical scenarios, with higher temperatures and higher neutrino chemical potentials, allow  higher positive values of the strangelet's  electric charge at the boundary and consequently smaller values of $\sigma$. {In contrast, $\gamma$ is strongly dominated by the density of light ($u$ and $d$) quarks and is quite independent of the charge-per-baryon ratio, the temperature and neutrino trapping. We discuss the relative importance of surface and curvature effects as well as} some astrophysical consequences of these results.

\end{abstract}

\maketitle

\section{Introduction}

It has been conjectured some decades ago that the so called strange quark matter (SQM), made up of roughly the same amounts of up, down and strange quarks, could be the true ground state of strongly interacting matter \cite{Itoh:1970uw, Bodmer:1971we, Witten:1984rs, Farhi:1984qu}. Such possibility attracted much interest in the 1980s (see the papers in \cite{Aarhus1991} and references therein) and several applications of SQM in physics and astrophysics were investigated, including the possible existence of \textit{strangelets} (tiny quark lumps of SQM)  and \textit{strange stars} (compact stars where SQM extends from the stellar center all the way to the stellar surface). 

The interface between quarks and the vacuum is of crucial importance for understanding the properties of stellar objects made of SQM.  In the pioneer study of Alcock, Farhi and Olinto \cite{Alcock:1986hz},  it was suggested that strange stars would be characterized by an enormous density gradient at the surface and may have a thin crust with the same composition as the pre-neutron drip outer layer of a conventional neutron star crust, supported by a large outward-directed electric field due to quarks. This possibility was reexamined more recently by Jaikumar, Reddy and Steiner \cite{Jaikumar:2005ne} who suggested that  a homogeneous and locally charged neutral phase of quarks and electrons could become unstable to phase separation at small pressures. Instead, a mixed phase made up of strangelets embedded in a uniform electron background would be energetically favored if the quark matter surface tension is below some critical value.  
The stability of these strangelets was further analysed in detail taking into account electrostatic effects, including Debye screening, and an arbitrary surface tension at the interface between vacuum and quark matter \cite{Alford:2006bx}. It was found that below a critical surface tension of the order of a few $\mathrm{MeV} / \mathrm{fm}^{2}$, large strangelets are unstable to fragmentation and the surface of a strange star will form a crust consisting of a crystal of charged strangelets immersed in a neutralizing background of electrons. At the surface of this crust there would be no electric field. The width of this strangelet-crystal crust was then calculated by Alford and Eby \cite{Alford:2008ge}. They showed that for a star of radius $10 \, \mathrm{km}$ and mass $1.5 M_{\odot}$ we can expect a thickness ranging from zero to hundreds of meters. When the strange quark is heavier and the surface tension is smaller the crust tends to be larger.  For smaller quark stars it will be even thicker.

As emphasized by Alford, Han and Reddy \cite{Alford:2011ue}, in the scenario of low surface tension strange stars are not self-bound compact objects, i.e. the strangelet crust is bound to the star gravitationally and not by strong interactions as in traditional strange stars. In this sense, they  resemble stars made of nuclear matter and the mass–radius relation has a branch of compact stars and another branch of low-mass large-radius objects (analogous to white dwarfs)  denominated strangelet dwarfs \cite{Alford:2011ue}.

In this work we will study the surface tension {and the curvature coefficient} of strangelets that would compose the strangelet crust of strange stars or the interior of strangelet dwarfs described before. 
In large strangelets, Debye screening causes the positive charge density to migrate towards the surface, resulting in a charged skin whose thickness is of order the Debye length $\lambda_{D}$ ($\sim 5 \,  \mathrm{fm}$ in quark matter) and a neutral interior \cite{Alford:2006bx}. In contrast, in small enough strangelets electric charge will be distributed all along their interiors. However, the surface tension is determined by the thermodynamic state within a \textit{thin layer} below the strangelet boundary. The thickness of this layer must be of the order of the strong interaction's length scale, therefore it is extremely thin ($\ll \lambda_{D}$). To let the analysis as general as possible,  we will not describe in this work the internal structure of these strangelets within a self-consistent formalism but rather calculate the surface tension taking as inputs the values of the relevant thermodynamic quantities at the strangelet boundary. We will assume that strange quark matter is a mixture of $u$, $d$ and $s$ quarks described within the MIT bag model plus  electrons and neutrinos, all them in chemical equilibrium under weak interactions. We will adopt typical thermodynamic conditions prevailing in cold strange stars, hot leptonized proto strange stars and hot post binary merger strange stars.  Finite size effects will be included by means of the multiple reflection expansion (MRE) formalism  \cite{Balian:1970fw, Madsen:1994vp, Kiriyama:2002xy, Kiriyama:2005eh, Lugones:2011xv, Lugones:2015bya, Lugones:2013ema}. 

This paper is organized as follows. In Sec. \ref{sec:surface_tension} we summarize the MRE formalism and describe the thermodynamic constrains imposed on strange quark matter such as chemical equilibrium, neutrino trapping and global electric charge neutrality. In Sec. \ref{sec:results} we present our results and in Sec. \ref{sec:conclusions} our main conclusions.

\section{Surface tension of charged strangelets}
\label{sec:surface_tension}

\subsection{The model}

Let us consider an electrically charged strangelet  composed by $u$, $d$, and $s$ quarks, electrons, and neutrinos, all in chemical equilibrium under weak interactions. Surface tension is determined by the state of strange quark matter at the strangelet's boundary, thus, we will focus on particles in that region. The single-particle dispersion relation for quarks and electrons has the form: 
\begin{equation} 
E = (k^2 + m^2)^{1/2}  + q |e \phi|
\end{equation}
where $\phi$ is the electric potential at the boundary, $m$ is the particle's mass, $k$ its momentum, $E$ its energy, and we have $q_u=2/3$, $q_d=-1/3$, $q_s=-1/3$, $q_e=-1$ for quarks $u$, $d$, $s$ and electrons respectively.
In general, the electric potential is a function of the position inside the strangelet and its surface value $\phi$ should be determined self-consistently by means of the Poisson equation (see e.g. Ref. \cite{Alford:2006bx}). However, to keep our analysis as general as possible, we will  treat $\phi$ as an input parameter in this work. 

Effects due to the finite size of the strangelet will be taken into account within the MRE framework  \cite{Balian:1970fw,Madsen:1994vp,Kiriyama:2002xy,Kiriyama:2005eh}.  In this formalism, the  modified  density of states of a finite spherical droplet is given by: {
\begin{equation}
\rho_{\mathrm{MRE},i}  \left(k, m_i, R\right)=1+\frac{6 \pi^{2}}{k R} f_{S,i}+\frac{12 \pi^{2}}{(k R)^{2}} f_{C,i} ,
\label{rho_MRE}
 \end{equation}
where 
\begin{equation}
f_{S,i}(k) = - \frac{1}{8 \pi} \left(1 - \frac{2}{\pi} \arctan \frac{k}{m_i} \right) 
\label{eq:fs}
\end{equation}
is  the surface contribution to the new density of states, and the curvature contribution is given in the Madsen ansatz \cite{Madsen:1994vp}
\begin{equation}
f_{C,i}(k)=\frac{1}{12 \pi^{2}}\left[1-\frac{3 k}{2 m_i}\left(\frac{\pi}{2}-\arctan \frac{k}{m_i}\right)\right].
\label{eq:fc}
\end{equation}
}

When finite size effects are added, the thermodynamic integrals are obtained from the bulk ones by means of the following replacement \cite{Lugones:2018qgu,Lugones:2016ytl}:
\begin{equation}
\int_{0}^{\infty} \cdots \frac{k^{2} d k}{2 \pi^{2}} \longrightarrow \int_{\Lambda_{\mathrm{IR}}}^{\infty} \cdots \frac{k^{2} d k}{2 \pi^{2}} \rho_{\mathrm{MRE}}
\label{eq:prescription}
\end{equation}
where $\Lambda_{\mathrm{IR}}$ is the largest solution of the equation $\rho_{\mathrm{MRE}}(k)=0$ with respect to the momentum $k$.  
$\Lambda_{\mathrm{IR}}$ depends on $m$ and $R$ and its values are given in Table \ref{table:cutoff}. 
{Since leptons form a uniform background, the above prescription applies only to quarks, which feel the strong interaction and are confined within a finite spherical region.}

\begin{table}[tb]
\centering
\begin{tabular}{c c c c c }
\hline \hline
particles & $m \, \mathrm{[MeV]}$   & $R \, \mathrm{[fm]}$   & $\Lambda_{\mathrm{IR}} \, \mathrm{[MeV]}$    \\
\hline  
quarks u, d & 5     &   3     & 51.33  \\ 
            & 5     &   5     & 32.47 \\   
            & 5     &  10    & 18.02 \\  

\hline
quarks s &150     &   3    & 96.89 \\
         &150     &   5    & 63.00 \\   
         &150     &  10    & 33.65 \\ 
\hline \hline 
\end{tabular}
\caption{Infrared cutoff $\Lambda_{\mathrm{IR}}$ for different particle masses $m$ and different values of the drop's radius $R$.} 
\label{table:cutoff}
\end{table}

{We will describe quark matter by means of the grand thermodynamic potential of the MIT bag model including finite size effects:
\begin{equation}
\begin{aligned}
\Omega =& - \sum_{i=u, d, s} \frac{g_i V}{6 \pi^{2}} \int_{\Lambda_{\mathrm{IR},i}}^{\infty} \frac{\partial E_i}{\partial k} \left(F_{i} + \bar{F}_{i} \right)  \rho_{\mathrm{MRE},i}    k^3 d k \\
   & + B V - \sum_{i=e, \nu_e} \frac{g_i V}{6 \pi^{2}} \int_{0}^{\infty}  \frac{(F_i + \bar{F}_i)}{\sqrt{k^2 + m_i ^2}}  k^4 d k, 
\end{aligned}    
\label{eq:Omega}
\end{equation}
where $B$ is the bag constant, and $g_i$ is the particle's degeneracy ($6$ for quarks, $2$ for electrons and $1$ for neutrinos). The Fermi--Dirac distribution functions for particles and antiparticles are:
\begin{equation}
\begin{aligned} F_{i} &=\frac{1}{e^{\left(E_{i}-\mu_{i}\right) / T}+1}, \\ 
\bar{F}_{i} &=\frac{1}{e^{\left(E_{i}+\mu_{i}\right) / T}+1}, 
\end{aligned}
\end{equation}
where $\mu_i$ is the chemical potential. Eq. \eqref{eq:Omega} can be rearranged as: 
\begin{equation}
\Omega = - P V + \sigma S + \gamma C ,
\end{equation}
where the pressure $P$, the surface tension $\sigma$ and the curvature coefficient $\gamma$ can be obtained respectively from the volume ($V=\tfrac{4}{3}\pi R^3$), the surface ($S=4\pi R ^2$), and the curvature ($C=8 \pi R$) contribution to the thermodynamic potential (see \cite{Lugones:2018qgu, Lugones:2016ytl} for further details):
\begin{eqnarray}
\begin{aligned}
P =&  \sum_{u,d,s} \frac{g_i}{6 \pi^{2}} \int_{\Lambda_{\mathrm{IR},i}}^{\infty} \frac{(F_i + \bar{F}_i)}{\sqrt{k^2 + m_i ^2}}  k^4 d k - B  \\
&  + \sum_{e, \nu_e} \frac{g_i}{6 \pi^{2}} \int_{0}^{\infty}  \frac{(F_i + \bar{F}_i)}{\sqrt{k^2 + m_i ^2}}  k^4 d k, 
\end{aligned}
\end{eqnarray}
\begin{eqnarray}
\sigma = \sum_{u,d,s} \sigma_i = \sum_{u,d,s}  \frac{g_i}{3} \int_{\Lambda_{\mathrm{IR},i}}^{\infty}   \frac{(F_{i} + \bar{F}_{i}) f_{S, i} k^3 d k }{\sqrt{k^2 + m_i^2}} , 
\label{eq:sigma}
\end{eqnarray}
\begin{eqnarray}
\gamma = \sum_{u,d,s} \gamma_i = \sum_{u,d,s}  \frac{g_i}{3} \int_{\Lambda_{\mathrm{IR},i}}^{\infty}   \frac{(F_{i} + \bar{F}_{i}) f_{C, i} k^2 d k }{\sqrt{k^2 + m_i^2}} . 
\label{eq:gamma_integral}
\end{eqnarray}
Notice that  $B$ is absorbed in the pressure $P$ and therefore, the surface tension and the curvature coefficient are independent of the bag constant in the present model. Since we focus here only on $\sigma$ and $\gamma$, we will not adopt any specific $B$, but we will assume that its value is such that the energy per baryon of bulk quark matter at vanishing pressure and temperature is lower than the neutron's mass (the so called Bodmer-Witten-Terazawa conjecture \cite{Bodmer:1971we,Witten:1984rs}).
}

{
The particle number density  $n_i$ for the $i\mathrm{th}$ quark species is:
\begin{equation}
n_{i}=    \frac{g_i}{2 \pi^2}  \int_{\Lambda_{\mathrm{IR},i}}^{\infty} \left(F_{i} - \bar{F}_{i} \right) \rho_{\mathrm{MRE},i}  
 k^2 dk ,
\label{eq:number_density}
\end{equation}
and for electrons and electron neutrinos we have:  
\begin{equation}
n_{i}=    \frac{g_i}{2 \pi^2}  \int_{0}^{\infty} \left(F_{i} - \bar{F}_{i} \right)   k^2 dk, 
\end{equation}
with the dispersion relations $E_e = (k^2 + m_e^2)^{1/2}  - |e \phi|$ and $E_{\nu} = k$  respectively. }

\subsection{Thermodynamic constrains}

In this work we are interested in SQM nuggets immersed in a leptonic background at the crust of strange stars or the interior of strangelet dwarfs. In this situation quark matter is in chemical equilibrium under weak interactions, which means that the chemical potentials of different species are related by 
\begin{eqnarray}
\mu_d &=& \mu_u + \mu_e - \mu_{\nu_e},  \label{eq:chenical_1} \\
\mu_s &=& \mu_d  .    \label{eq:chenical_2}
\end{eqnarray}
Additionally, charge neutrality should be imposed globally (not locally). In this work, we will focus on the electric charge of quark matter near the strangelet's boundary because $\sigma$ is calculated there. We will write the global charge neutrality condition in a form that expresses the fact that matter at the boundary can be locally charged: 
\begin{eqnarray}
\left( \tfrac{2}{3} n_u - \tfrac{1}{3} n_d - \tfrac{1}{3} n_s - n_e\right)  = n_Q \ge  0.
\end{eqnarray}
The charge density $n_Q$ at the boundary must be positive because the lepton background is composed by negative particles (in our model only electrons).
Similarly to $\phi$, the value of $n_Q$ should be determined self consistently using the Poisson equation for each specific strangelet. In fact, this is the approach adopted in Ref. \cite{Alford:2006bx} where explicit profiles of these quantities are obtained.  However, notice that in the calculations of Ref. \cite{Alford:2006bx} the surface tension enters as a free parameter. In our work we will adopt a different (and in some sense complementary) approach. We will treat $n_Q$ as a free input quantity in order to determine the sensitivity of the surface tension to changes in this variable. 
For convenience, we will write the charge density in terms of the charge-per-baryon ratio:  
\begin{equation}
\xi \equiv \frac{n_Q}{n_B} ,   
\end{equation}
and will span all possible values of $\xi$. Therefore, our condition for the electric charge density will read:
\begin{eqnarray}
\tfrac{2}{3} n_u - \tfrac{1}{3} n_d - \tfrac{1}{3} n_s - n_e = \xi n_B .
\label{eq:finite_charge}
\end{eqnarray}
{Notice that, unlike Ref. \cite{Alford:2006bx}, our focus here is not to determine the mechanical and chemical equilibrium between both sides of a drop's interface, but to study the microscopic behavior of the thermodynamic coefficients $\sigma$ and $\gamma$. Because of this reason, electrostatic corrections to the pressure and the energy density don't play a role in our calculations, although they are certainly important for determining self consistently the size and internal profiles of strangelets.}

\subsection{Astrophysical scenarios}
\label{sec:2c}

We consider here three different astrophysical scenarios characterized by the temperature as well as by the amount of  trapped neutrinos in the strange star:

\begin{itemize}

\item \textit{Cold deleptonized strange stars (CSS)}. This is the case of most strange stars a few minutes after their birth. The thermodynamic state can be characterized by a very low temperature (typically below $1 \, \mathrm{MeV}$) and no trapped neutrinos because their mean free path is much larger that the stellar radius. As a representative case, we shall consider here $T=1 \, \mathrm{MeV}$  and $\mu_{\nu_e}=0$.

\item  \textit{Hot lepton rich proto strange stars (PSS)}.  This would be the case during the first few  minutes after their birth \cite{Pons:2001ar}. Matter is at high temperatures (typically up to $\sim 40 \, \mathrm{MeV}$) and there is a large amount of trapped neutrinos in the system (neutrino chemical potential $\mu_{\nu_e}$ up to $\sim 150 \, \mathrm{MeV}$). As a representative case we consider here $T= 30 \, \mathrm{MeV}$ and $\mu_{\nu_e} = 100 \, \mathrm{MeV}$. 

\item \textit{Post merger strange stars (PMSS)}.  The LIGO/Virgo collaboration has recently detected the first signal of gravitational waves coming from the binary neutron star merger GW170817 \cite{TheLIGOScientific:2017qsa}. Numerical simulations of these events that include quark matter cores suggest that the post merger object may attain very high temperatures (several tens of $\mathrm{MeV}$  \cite{Most:2018eaw}) and contain a huge amount of trapped neutrinos. To the best of our knowledge, simulations for the specific case of a post merger object made up of self-bound strange quark matter are not available yet. As a limiting case we consider here $T= 100$ MeV and $\mu_{\nu_e} = 200$ MeV. 

\end{itemize}

\begin{figure*}[tb]
\centering
\includegraphics[angle=0,scale=0.33]{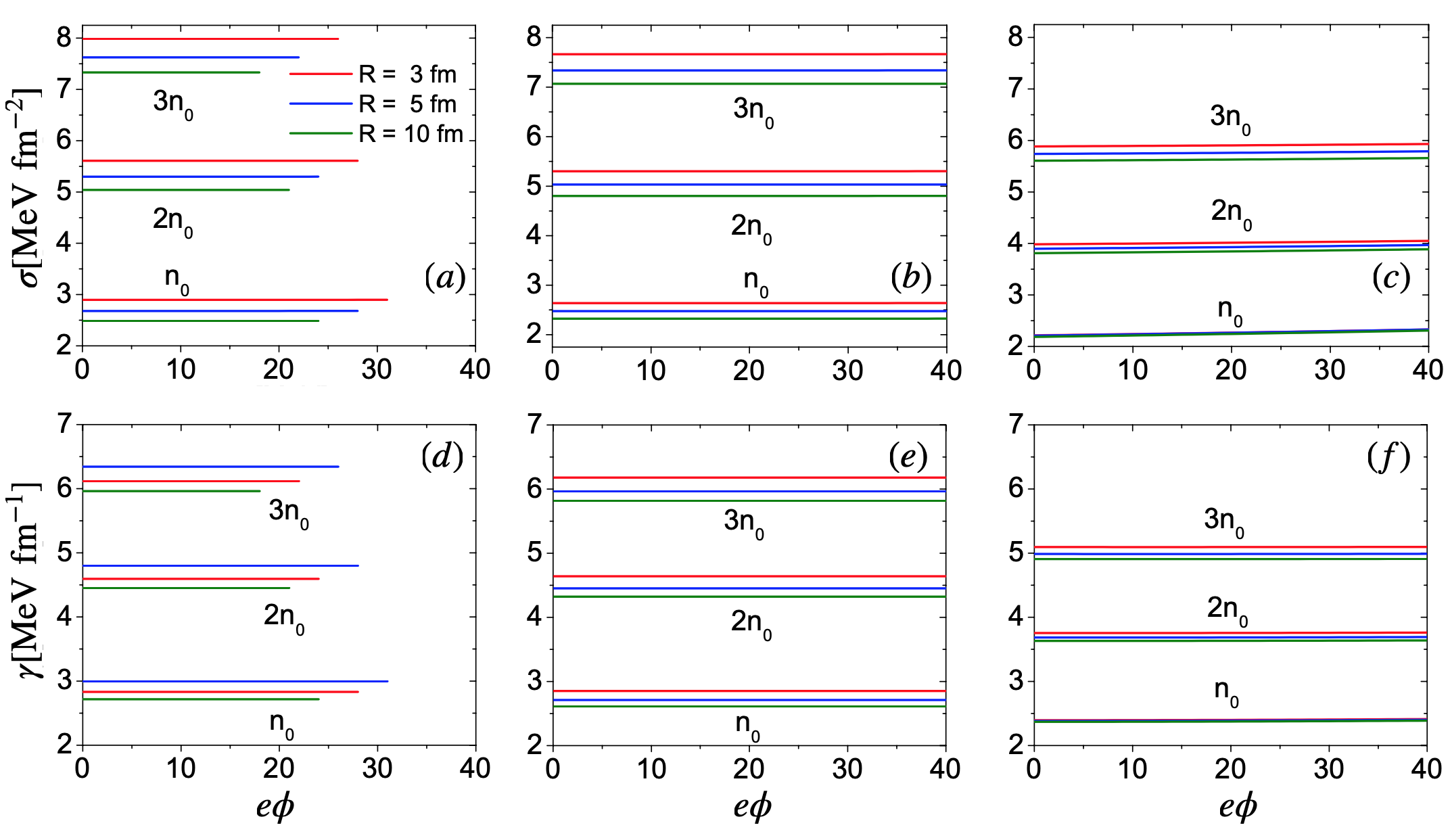}
\caption{Surface tension {and curvature coefficient} of strangelets as a function of the electric potential $\phi$ at the strangelet's boundary for:  cold strange stars {(panels $a$ and $d$)}, hot lepton-rich proto strange stars {(panels $b$ and $e$)} and  strange stars formed after a binary merger event {(panels $c$ and $f$)}. We show three groups of curves with baryon number densities $n_B = n_0, 2 n_0, 3 n_0$.  }
\label{phi}
\end{figure*}

\section{Results and discussion}
\label{sec:results}

\begin{figure*}[tb]
\centering
\includegraphics[angle=0,scale=0.33]{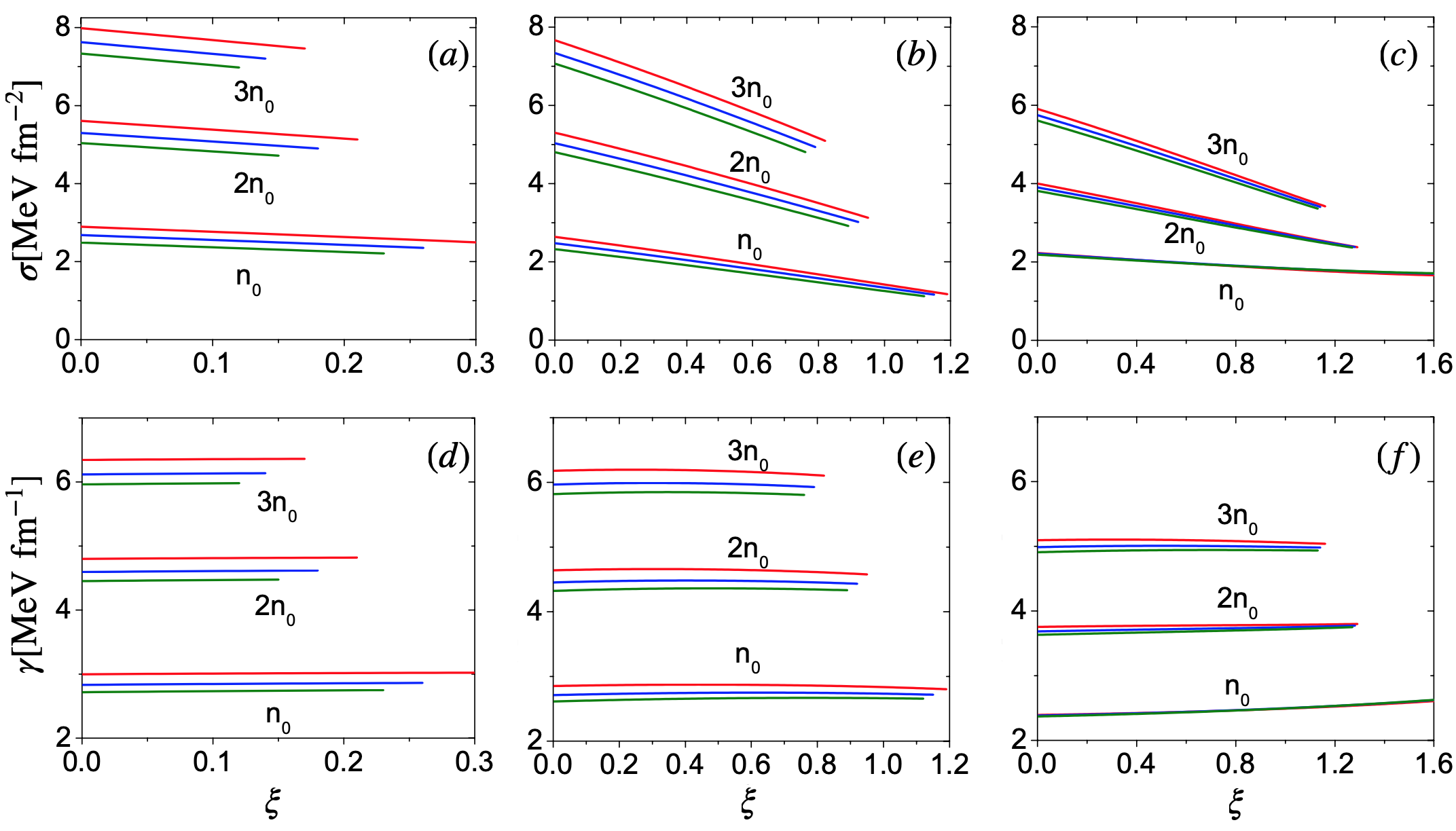}
\caption{Surface tension {and curvature coefficient} as a function of the charge-per-baryon ratio $\xi$ for the same astrophysical scenarios and the same densities as in previous figure. Notice that the range of $\xi$ is different in each panel. Colors have the same meaning as in Fig. \ref{phi}. }
\label{X}
\end{figure*}

\begin{figure*}[tb]
\centering
\includegraphics[angle=0,scale=0.33]{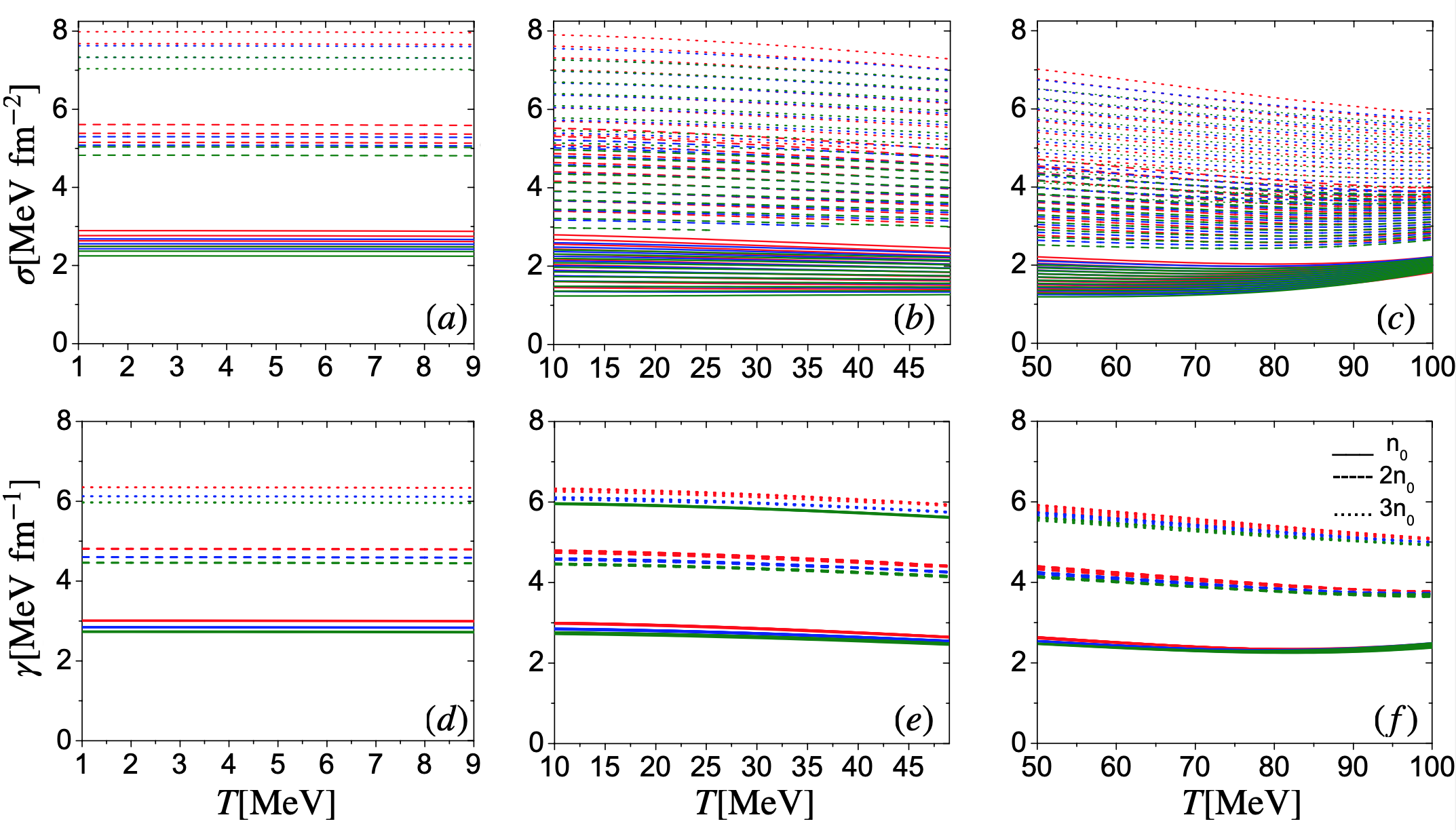}
\caption{Surface tension {and curvature coefficient} of strangelets as a function of the temperature. We spanned a different range of $T$ for each astrophysical scenario, namely $1 \, \mathrm{MeV} < T < 9 \, \mathrm{MeV}$ for CSSs (panel $a$), $10 \, \mathrm{MeV} < T < 49 \, \mathrm{MeV}$   for PSSs (panel $b$) and  $50 \, \mathrm{MeV} < T < 100 \, \mathrm{MeV}$ for PMSSs (panel $c$). For each density, the upper curve has $\xi=0$ (local charge neutrality), and the curves below it have $\xi$ that increases in steps of $0.1$ (charged strangelets).}
\label{T}
\end{figure*}

\begin{figure*}[tb]
\centering
\includegraphics[angle=0,scale=0.33]{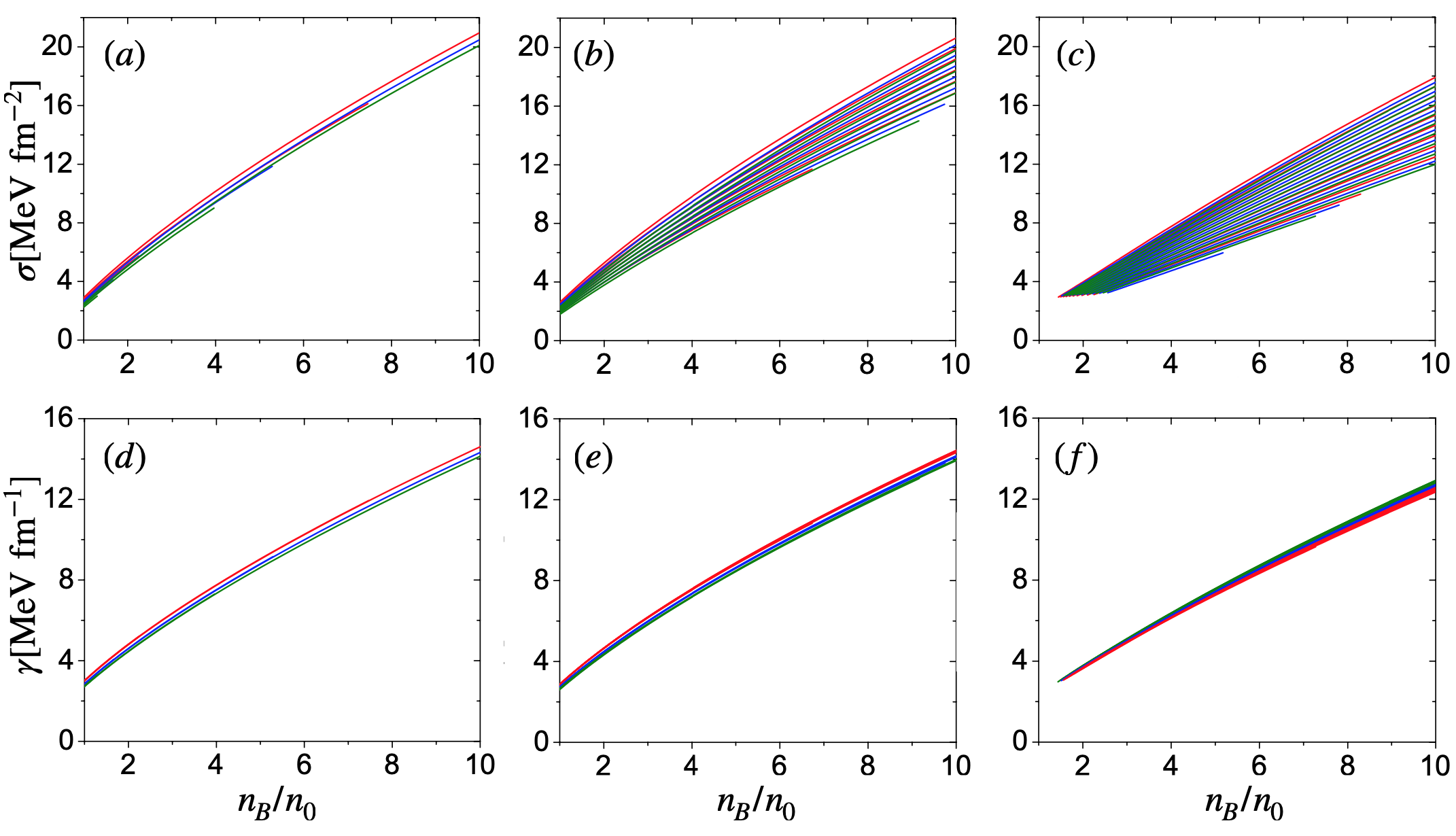}
\caption{Surface tension {and curvature coefficient} as a function of the baryon number density at the strangelet's boundary for the same astrophysical scenarios of previous figures. The upper curvue has $\xi=0$ (local charge neutrality), and the curves below it have $\xi$ that increases in steps of $0.1$ (charged strangelets).  Note that the axes in panel $c$ have a different scale. }
\label{Bdensity}
\end{figure*}

\begin{figure*}[tb]
\centering
\includegraphics[angle=0,scale=0.34]{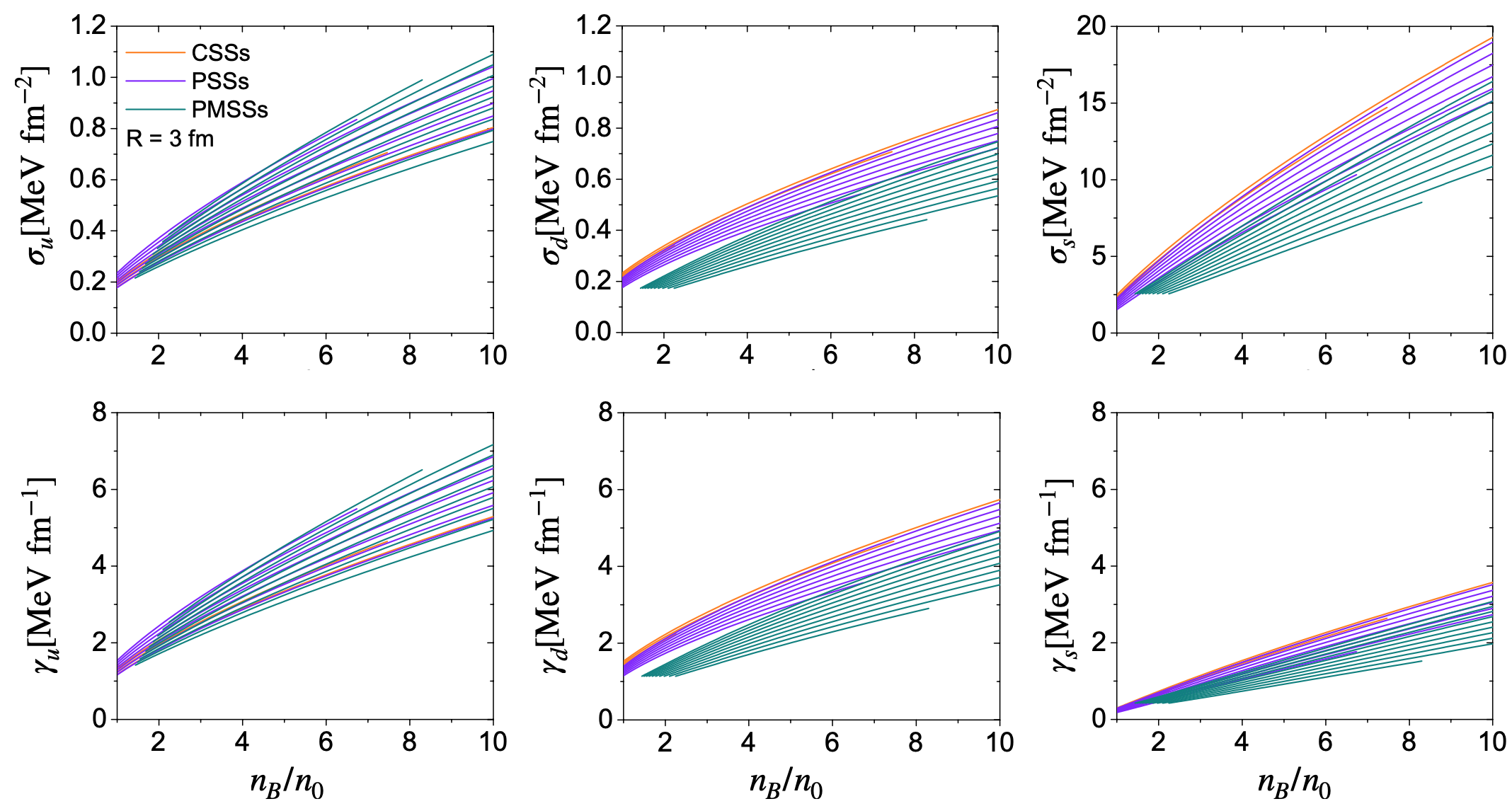}
\caption{{Contribution of each flavor to the surface tension and the curvature coefficient. Differently from previous figures, each panel contains results for the three astrophysical scenarios of Sec. \ref{sec:2c}. } }
\label{fig:flavors}
\end{figure*}

\begin{figure*}[tb]
\centering
\includegraphics[angle=0,scale=0.33]{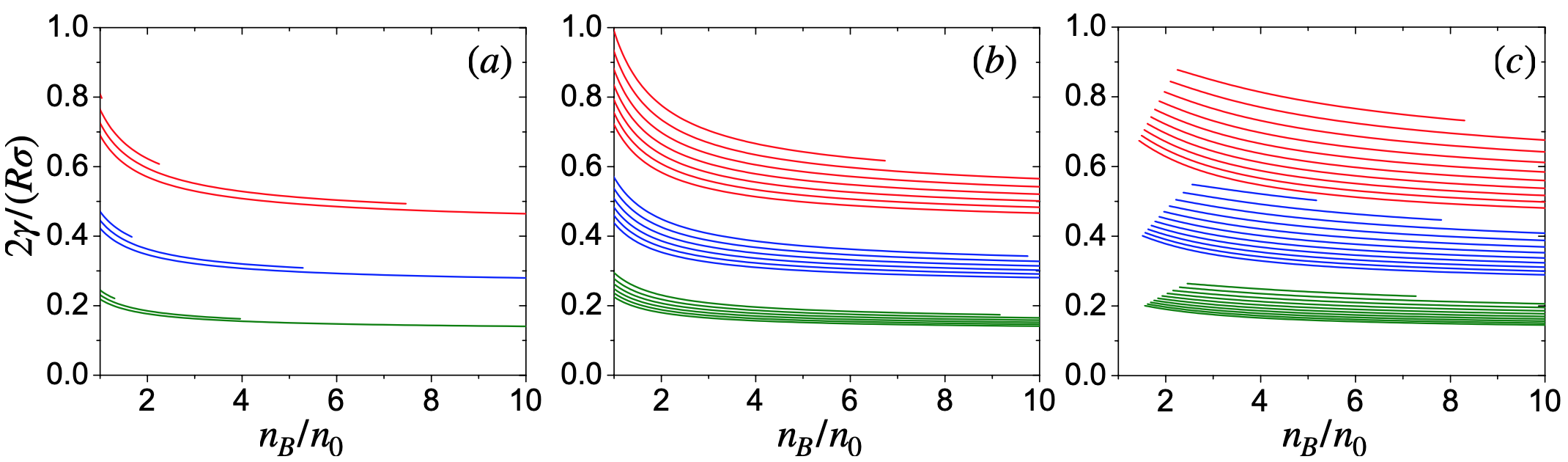}
\caption{{Ratio of the curvature and the surface terms in the grand thermodynamic potential. Each panel corresponds to an astrophysical scenario of Sec. \ref{sec:2c}: (a) CSSs, (b) PSSs, and (c) PMSSs. Colors have the same meaning as in Fig. \ref{phi}. For each color,  the lowest curve has $\xi=0$ (local charge neutrality), and the curves above it have $\xi$ that increases in steps of $0.1$ (charged strangelets).}  }
\label{fig:ratio}
\end{figure*}

We present here our results for the surface tension $\sigma$ {and the curvature coefficient $\gamma$}, obtained by means of Eqs. \eqref{eq:sigma} - \eqref{eq:gamma_integral} and supplemented by the conditions of chemical equilibrium (Eqs. \eqref{eq:chenical_1} and \eqref{eq:chenical_2}) and finite electric charge (Eq. \eqref{eq:finite_charge}). As explained in previous sections, $\sigma$ {and $\gamma$ are} determined by the state of matter at the strangelet's boundary. In order to obtain the boundary's state, the profile of all thermodynamic quantities inside the strangelet should be determined self-consistently by imposing global charge neutrality and solving the Poisson equation for a strangelet surrounded by electrons located within a Wigner-Seitz cell (see e.g. Ref. \cite{Alford:2006bx}). However, in this work we adopt a different approach.  The boundary values of the electric potential $\phi$, the baryon number density $n_B$, the charge-per-baryon ratio $\xi$, the neutrino chemical potential $\mu_{\nu_e}$ and the temperature $T$ will be treated as input parameters, as well as the strangelet radius $R$. Since strangelets with radii much larger than $\lambda_D$ are not expected to form,  we will consider $R = 3 \, \mathrm{fm}$, $5 \, \mathrm{fm}$ and $10 \, \mathrm{fm}$, which are of the order of magnitude of $\lambda_D$. The values of $T$ and $\mu_{\nu_e}$ are fixed according to the three different astrophysical scenarios presented before.

In Fig. \ref{phi} we show $\sigma$ {and $\gamma$ as functions} of the boundary's electric potential $\phi$.  According to Refs. \cite{Alford:2006bx,Alford:2008ge}, the electric field is zero at the strangelet's center and grows roughly linearly up to a few $\mathrm{MeV} / \mathrm{fm}$ at the strangelet's boundary. For $R \sim \lambda_D \sim 5 \mathrm{fm}$ and a surface electric field of $4 \, \mathrm{MeV} / \mathrm{fm}$ (see \cite{Alford:2006bx,Alford:2008ge}) we find that the electric potential $\phi$ at the boundary is around $20 \, \mathrm{MeV}$. In Fig. \ref{phi} we let $\phi$ vary up to  $50 \, \mathrm{MeV}$. The three different groups of lines correspond to different baryon number densities at the boundary, namely $n_0$, $2 n_0$ and $3 n_0$ (being $n_0$ the nuclear saturation density). {Both, $\sigma$ and $\gamma$, change} significantly with density but are quite insensitive to $\phi$,  in the three astrophysical scenarios considered here.  Due to that fact we will adopt $\phi =0$  in the rest of our numerical calculations. 
Notice that some curves do not span the whole range of the dependent variable. This happens because the electron chemical potential goes to zero as one approaches to the right end of the curve. Solutions beyond this point would have negative electron chemical potential corresponding to an excess of positrons inside the strangelet, thus increasing its positive charge. Such possibilities are not relevant in our case. This behavior is present in all figures shown here.

In Fig. \ref{X} we show the behavior of the surface tension {and the curvature coefficient}, with the charge-per-baryon ratio $\xi$ for the three values of the density considered before. Similarly to the previous figure, $\sigma$ {and $\gamma$} increase significantly with the density for a fixed value of $\xi$.  Within each group of constant density, $\sigma$ {and $\gamma$} get larger as the size becomes smaller. At low temperatures and high densities, the strangelet size has a non-negligible role in $\sigma$ {and $\gamma$}, but the curves for different radii get closer as the $T$ grows and/or $n_B$ decreases. 
In the CSS case (panel $a$), we see that $\sigma$ slightly decreases with $\xi$ for all the densities considered. In the PSS and PMSS cases (panels $b$ and $c$), we observe a more significant decline of $\sigma$ as the positive electric charge at the strangelet's boundary becomes greater. This behavior can be understood as follows. The value of $\sigma$ is dominated by the contribution of $s$ quarks, because they are the most massive particles in the system. If the strangelet is forced to have a large positive electric charge (large $\xi$), negatively charged $s$-quarks will be suppressed in the system and their contribution to $\sigma$ will diminish, resulting in smaller values of the surface tension. {On the hand, $\gamma$ depends mainly on the $u$ and $d$ quarks contribution (see below) and is quite insensitive to $\xi$ in all astrophysical scenarios.}

In Fig. \ref{T} we show {$\sigma$ and $\gamma$} as functions of $T$ for three values of the density $n_0$, 2$n_0$ and 3$n_0$ (solid, dashed and dotted lines respectively). For a given value of $T$, { both $\sigma$ and $\gamma$} grow with the density and decrease for larger drops. In CSSs, the surface tension {and the curvature coefficient} are insensitive to the temperature. In PSSs, there is a small decrease {of both $\sigma$ and $\gamma$} with $T$ for the highest density considered here ($3 n_0$). For PMSSs, both $\sigma$ {and $\gamma$} decrease with $T$ for the cases of $2n_0$ and $3n_0$.
As seen in previous figures, we notice that strangelets with larger positive charges are allowed in the most extreme astrophysical scenarios ($\xi$ may have larger values in panels $b$ and $c$).

In Fig. \ref{Bdensity} we display {$\sigma$ and $\gamma$} as functions of $n_B/n_0$ for the same scenarios as in previous figures. In all cases we find that the surface tension {and the curvature coefficient} considerably grow with increasing $n_B$, {with a roughly linear dependence}. At low temperatures large values of $\xi$ are not allowed because they would lead to $\mu_e < 0$.

{We display in Fig. \ref{fig:flavors} each flavor contribution to the surface tension and curvature, as a function of $n_B/n_0$, for the specific case with $R=3$ fm and the three astrophysical scenarios of Sec. \ref{sec:2c}. Notice that $\sigma_s$ (panel $c$) is one order of magnitude larger than $\sigma_u$ and $\sigma_d$ (panels $a$ and $b$) while $\gamma_u$ and $\gamma_d$ (panels $d$ and $e$) are roughly twice $\gamma_s$ (panel $f$). 
This behavior can be understood if one keeps in mind that for massless particles $f_S =0$ and  $f_C = -1/(24 \pi^2)$ (as shown in  \cite{Madsen:1994vp}) while for $m \rightarrow \infty$ both $f_C$ and $f_S$ remain finite. This means that light quarks tend to have a minor contribution to the surface tension, but have a dominant role in the curvature contribution. On the other hand, both effects, surface and curvature, are relevant in principle for massive enough species.  
This can be seen in Fig. \ref{fig:ratio} where we show the ratio $\gamma C /(\sigma S) = 2 \gamma /(R \sigma)$ between surface and curvature terms in the grand thermodynamic potential as way of assessing the relative influence of each contribution. In general, the curvature contribution is significant in all cases but becomes more relevant for smaller drop sizes. For typical densities above $2 n_0$ the ratio is roughly constant. For $R=10$ fm, the curvature term has a contribution of about  $20\%$ with respect to the surface one. For $R=3$ fm, this contribution grows above $50 \%$ for CSSs and is even larger ($70-80 \%$) in the PMSSs case. For lower densities the ratio increases even more. }

\section{Summary and Conclusions}
\label{sec:conclusions}

In the present work we have studied the surface tension {and the curvature coefficient} of positively charged strangelets in global electric charge equilibrium with an electron background.  Strange quark matter composed by $u, d$ and $s$ quarks, electrons and neutrinos in chemical equilibrium under weak interactions was described within the MIT bag model. {Electrons and neutrinos do not contribute explicitly to $\sigma$ and $\gamma$ because quarks are embedded in a uniform leptonic background (we assume that there is no boundary for leptons). However, electrons have a significant indirect role on $\sigma$ and $\gamma$ because of their contribution to global charge neutrality and chemical equilibrium. Similarly, trapped neutrinos are relevant because they shift the chemical equilibrium conditions.}  Finite size effects were included within the MRE framework for strangelets with radii $3\,\mathrm{fm}$, $5\,\mathrm{fm}$ and $10\, \mathrm{fm}$ which are of the order of the Debye length in strange quark matter ($\lambda_D \sim 5 \, \mathrm{fm}$).

The surface tension {and the curvature coefficient are} determined by the state of quark matter within a thin layer below the strangelet's  boundary.  To let the analysis as general as possible,  we didn't determine the internal structure of strangelets but calculated $\sigma$ {and $\gamma$} taking as inputs the boundary values of the electric potential $\phi$, the baryon number density $n_B$, the charge-per-baryon ratio $\xi$, the neutrino chemical potential $\mu_{\nu_e}$ and the temperature $T$, as well as the strangelet radius $R$. 
For typical values of the electric potential at the strangelet's boundary \cite{Alford:2006bx,Alford:2008ge},  our results show that $\sigma$ {and $\gamma$ are} quite insensitive to $\phi$ in all astrophysical scenarios.  
We also explored the dependence of the surface tension {and the curvature coefficient} with the charge-per-baryon ratio $\xi$. In general, $\sigma$  decreases  as the strangelet's boundary becomes more positively charged.  This effect may be large in some cases; for example in the PSS and PMSS cases, $\sigma$ may be reduced by a significant factor when comparing a charge neutral strangelet ($\xi = 0$) with  maximally charged strangelets ($\xi > 1$) at $n_B = 1 \, n_0$.
This occurs because $\sigma$ is dominated by the contribution of $s$ quarks which are the most massive particles in the system. Negatively charged $s$-quarks are suppressed in strangelets with a large positive electric charge (large $\xi$),  diminishing their contribution to $\sigma$ and resulting in smaller values of the total surface tension. 
We also find that smaller strangelets admit slightly more positive electric charge than larger strangelets, and that $\sigma$ decreases a few percent when the radius increases from $3 \, \mathrm{fm}$ to $10 \,  \mathrm{fm}$. 
{On the contrary, the curvature coefficient has contributions from all three flavors, with a dominance of $u$ and $d$ quarks, which makes $\gamma$ quite insensitive to $\xi$ in all astrophysical scenarios.}

The surface tension {and the curvature coefficient} grow significantly with the baryon number density at the strangelet's boundary.  {The values for $\sigma$ range between $3 \, \mathrm{MeV}/\mathrm{fm}^2$ for $n_B = 1 n_0$ to $20 \, \mathrm{MeV}/\mathrm{fm}^2$ for $n_B = 10 n_0$ while $\gamma$ varies between $3 \, \mathrm{MeV}/\mathrm{fm}$ for $n_B = 1 n_0$ and $15 \, \mathrm{MeV}/\mathrm{fm}$ for $n_B = 10 n_0$.}
In this work we are interested mainly in strangelets within the crust which have presumably densities below $\sim 3 n_0$. For these densities, the surface tension does not exceed the value $\sim 8 \, \mathrm{MeV}/\mathrm{fm}^2$ in agreement with the early calculations of Ref. \cite{Berger:1986ps} {and $\gamma$ is below $6 \, \mathrm{MeV}/\mathrm{fm}$}.  We also verified that the most extreme astrophysical scenarios, with higher temperatures and higher neutrino chemical potentials, allow  higher positive values of the strangelet's  electric charge at the boundary and consequently smaller values of $\sigma$.

{In order to estimate the relative influence of surface and curvature effects on strangelets we explored the ratio between the surface and curvature contributions in the grand thermodynamic potential. We find that the curvature contribution is significant in all cases. For $R=10$ fm, the curvature term has a contribution of about  $20\%$ with respect to the surface one. For $R=3$ fm, this contribution grows above $50 \%$ for CSSs and is even larger ($70-80 \%$) in the PMSSs case. For lower densities the ratio increases even more. }

{Let us compare our results with previous works and explore some astrophysical consequences.} It has been shown within a model independent approach, that there exists a critical surface tension $\sigma_{\text {crit}}$ below which quark star surfaces will fragment into a crystalline crust made of charged strangelets immersed in an electron gas. {The critical value is obtained by requiring that the strangelet and the leptonic background are in chemical and mechanical equilibrium  and that the energy cost of surface and electrostatic contributions increase the overall energy with respect to the normal strange quark matter state.} 
The critical value is given by \cite{Alford:2006bx}:
\begin{equation}
\sigma_{\mathrm{crit}} = 0.1325 \frac{n_{Q,0}^{2}}{\sqrt{4 \pi \alpha} \chi_{Q}^{3 / 2}}
\end{equation}
where $\alpha=1 / 137$,  $n_{Q,0}$ is the charge density of quark matter at zero electric charge chemical potential and $\chi_{Q} $ is the electric charge susceptibility of quark matter.  If $\sigma < \sigma_{\text {crit}}$, then the energetically favored structure for the crust would be a strangelet crystal and not a simple sharp surface as assumed in the early works on strange stars. Additionally, low-mass large-radius objects analogous to white dwarfs, called strangelet dwarfs would be possible \cite{Alford:2011ue}.
Assuming noninteracting three-flavor quark matter to the lowest nontrivial order in $m_{s}$, we have \cite{Jaikumar:2005ne,Alford:2008ge}:
\begin{equation}
n_{Q,0} =\frac{m_{s}^{2} \mu}{2 \pi^{2}},  \qquad 
\chi_{Q} =\frac{2 \mu^{2}}{\pi^{2}} ,    
\end{equation}
from where one gets:
\begin{equation}
\sigma_{\mathrm{crit}} = 36\left(\frac{m_{s}}{150 \mathrm{MeV}}\right)^{3} \frac{m_{s}}{\mu} \mathrm{MeV} / \mathrm{fm}^{2} .
\end{equation}
Assuming $m_s =150 \, \mathrm{MeV}$ and $300 \, \mathrm{MeV} <\mu< 450 \, \mathrm{MeV}$ we obtain $\sigma_{\mathrm{crit}} \approx 12-18 \, \mathrm{MeV}/\mathrm{fm}^2$. {However, it is important to remark that the latter condition has been derived without considering the role of curvature effects. Since our results show that curvature effects are significant, the above equation  should be improved to include them. }

For the typical physical conditions considered in our calculations, the surface tension is comfortably below the critical value favoring the existence of a strangelet crust (cf. \cite{Jaikumar:2005ne}). Moreover, we have shown that low densities,  high temperatures, neutrino trapping and high enough electric charge at the strangelet's boundary decrease the surface tension {favoring values below $\sigma_{\mathrm{crit}}$. However, since curvature effects at densities around $\sim 1-3 \, n_0$ can be of the order of surface ones, such conclusion may be significantly altered.}
Additionally, notice that depending on the considered equation of state, $\sigma_{\mathrm{crit}}$ may vary significantly (see for example the results of Ref. \cite{Alford:2008ge} where  $\sigma_{\mathrm{crit}} = 0.5 - 12 \, \mathrm{MeV}/\mathrm{fm}^2$ is obtained). Thus, in a scenario of sufficiently small $\sigma_{\mathrm{crit}}$, our results would disfavor strangelet crusts and strangelet dwarfs. 

{Finally, it is worth mentioning that it has been recently conjectured that $u$ and $d$ quark matter can be more stable than strange quark matter \cite{Holdom:2017gdc}. Although such possibility is precluded within the standard MIT bag model, we can give some clues on the behavior of finite size effects on non-strange quark matter. In such a case the role of the surface tension is much less relevant that in the case of $uds$ matter because $\sigma_u$ and $\sigma_d$ are negligible. On the contrary, all finite size effects would arise from the curvature contribution, which would determine the existence (or not) of compact star crust containing absolutely stable $ud$ nuggets immersed in a leptonic background.}

\subsection*{Acknowledgements}
A. G. G. would like to acknowledge to CONICET for financial support under Grant No. PIP17-700. G. Lugones acknowledges the Brazilian agency Conselho Nacional de Desenvolvimento Cient\'{\i}fico e Tecnol\'ogico (CNPq) for financial support.

\bibliography{strangelets}

\end{document}